\documentclass[12pt]{article}

\usepackage{epsf}
\title{Causal loop quantum cosmology in momentum space}
\author{Ali Shojai, and Fatimah Shojai\\ Department of Physics, University of Tehran, Tehran, Iran.}
\date{}
\begin{document}
\maketitle
\begin{abstract}
We shall show that it is possible to make a causal interpretation of loop quantum cosmology using the momentum as the dynamical variable. We shall show that one can derive Bohmian trajectories. For a sample cosmological solution with cosmological constant, the trajectory is plotted.
\end{abstract}
\section{Introduction}
In order to introduce a definite trajectory for any quantum system, de-Broglie\cite{deb} and Bohm\cite{boh} presented the causal interpretation of quantum mechanics. It is proved that Copenhagen and causal interpretations are  identical for almost any statistical predictions of quantum theory, but there are predictions in the causal interpretation related to the trajectories which can not be asked in Copenhagen interpretation. This includes time of travel and time correlations and any other property having its origin in the particles trajectory. But the problem that whether such questions are physical or not is still unclear\cite{zei}. 

de-Broglie--Bohm theory is motivated by observing that the phase of the wave function of a non relativistic particle obeys the Hamilton--Jacobi equation modified by a potential called \textit{quantum potential}. 
This leads to the quantum trajectory. The power of this interpretation would be clear when one considers problems like measurement. It gives a deterministic version of measurement theory with the same statistical results as the Copenhagen interpretation.

Applying the theory to relativistic particles\cite{rel} and  fields\cite{boh1} is straightforward. It is also possible to make a causal interpretation of quantum gravity in terms of old variables\cite{old2} (i.e. using the components of the metric of spatial 3-surfaces in the ADM formalism as the dynamical variables). One can also make Bohmian loop quantum gravity\cite{ash} in the configuration space, that is assuming the connection as dynamical variables\cite{loop}.

Here we shall investigate the possibility of making a causal interpretation of loop quantum cosmology using the momentum as the dynamical variables. We shall first briefly review loop quantum cosmology and then de-Broglie--Bohm theory. Finally we shall show that although the momentum space of loop quantum cosmology is discrete, but it is possible to make a causal interpretation for it and derive the Bohmian trajectories.

\section{Loop quantum cosmology in a nutshell}
For a review of the subject, the reader is referred to \cite{boj}, but here we present a very short review of the main points. The classical phase space of a homogeneous isotropic cosmological model in terms of the ashtekar variables is given by $c$ and $p$, where 
\begin{equation}
c=\frac{1}{2}(k-\gamma\dot{a})
\end{equation}
and
\begin{equation}
|p|=a^2
\end{equation}
in which $a$ is the scale factor, $k$ is the curvature parameter and $\gamma$ is the Immirzi--Barbero parameter. The coordinate $c$ and momentum $p$ are normalized such that the Poisson bracket of them is:
\begin{equation}
\{c,p\}=\frac{16\pi G\gamma}{3}
\end{equation}
The Hilbert space of the quantized version of the theory is given by the eigenvalue problem of momentum:
\begin{equation}
\hat{p}|\ell\rangle=\frac{8\pi G\gamma\hbar}{3}\ell|\ell\rangle
\end{equation}
The gravitational Hamiltonian is given by 
\[
\hat{H}_g|\ell\rangle=
\]
\begin{equation}
\frac{3}{4\gamma^3\ell_0^3(16\pi G)^2\hbar}\left ( V_{\ell+\ell_0}-V_{\ell-\ell_0} \right ) \left ( e^{-i\ell_0k}|\ell+4\ell_0\rangle -\Omega |\ell\rangle +e^{i\ell_0k}|\ell-4\ell_0\rangle \right )
\end{equation}
in which $V_\ell$ are eigenvalues of the volume operator
\begin{equation}
V_\ell=\left ( \frac{8\pi G\gamma\hbar}{3}\ell\right )^{3/2}
\end{equation}
\begin{equation}
\Omega=2+\ell_0^2\gamma^2k
\end{equation}
and $\ell_0$ is a dimensionless parameter appeared in the theory because here we quantized the reduced model, instead of quantizing the full theory and then reducing the theory to our homogeneous isotropic model. There is an expectation that it is possible to obtain the value of $\ell_0$ by relating the full theory to the reduced one.

A general state of the system is 
\begin{equation}
|\psi\rangle=\sum_\ell\psi(\ell)|\ell\rangle
\end{equation}
For the homogeneous isotropic cosmological model the two gauge and diffeomorphism constraints are trivial and we only have to  apply the Hamiltonian constraint on the state:
\begin{equation}
\left ( \hat{H}_g+\hat{H}_m\right ) |\psi\rangle=0
\end{equation}
where $\hat{H}_m$ is the matter Hamiltonian. The above equation leads to:
\[
\left (V_{\ell+5\ell_0}-V_{\ell+3\ell_0}\right )e^{i\ell_0k}\psi(\ell+4\ell_0)- \Omega \left ( V_{\ell+\ell_0}-V_{\ell-\ell_0}\right )\psi(\ell) + 
\]
\begin{equation}
\left (V_{\ell-3\ell_0}-V_{\ell-5\ell_0}\right )e^{-i\ell_0k}\psi(\ell-4\ell_0)= -\frac{4}{3} \gamma^3\ell_0^3\hbar\left(16\pi G\right )^2H_m(\ell)\psi(\ell)
\end{equation}
Introducing a new function 
\begin{equation}
F(\ell)=\left (V_{\ell+\ell_0}-V_{\ell-\ell_0}\right )e^{ik\ell/4}\psi(\ell)
\label{e0}
\end{equation}
the Hamiltonian constraint reads as:
\[
F(\ell+4\ell_0)-\Omega F(\ell)+F(\ell-4\ell_0)=-\frac{4}{3} \gamma^3\ell_0^3\hbar\left(16\pi G\right )^2\frac{H_m(\ell)}{V_{\ell+\ell_0}-V_{\ell-\ell_0}}F(\ell)
\]
\begin{equation}
\equiv \hbar W(\ell) F(\ell)
\end{equation}
Although this is a difference equation, its argument is a continuous parameter. This difference equation can be decomposed into classes of difference equations with integer argument by changing the variable $\ell$ to:
\begin{equation}
\frac{\ell}{4\ell_0}=n+\epsilon
\end{equation}
where $n$ is integer and $\epsilon\in [0,1)$. In terms of $n$ the difference equation is:
\begin{equation}
F_\epsilon(n+1)-\Omega F_\epsilon(n)+ F_\epsilon(n-1)=\hbar W_\epsilon(n)F_\epsilon(n)
\label{AA5}
\end{equation}
Different values of $\epsilon$ defines different classes of difference equation. From now we shall drop the $\epsilon$ subscript for simplicity.
\section{The method of causal interpretation in a nutshell}
de-Broglie\cite{deb} and Bohm\cite{boh} have shown that it is possible to enlarge the theory of quantum mechanics such that one have a causal picture of the world. According to the causal theory of quantum mechanics, the state of a system is determined by $(q_i(t);\psi(q_i;t))$, in which $q_i(t)$ is the trajectory of the system and $\psi$ is the wave function. Bohm was able to show that one can choose a suitable equation of motion that firstly is consistent with the evolution of the wave function and secondly can lead to the correct prediction about the outcome of measurements in a causal manner. In other words the wave function reduction is not present but the quantum indeterminacy is coded in the initial values of $q_i$.

In order to illustrate the method of causal interpretation, let us focus on a non relativistic particle. The wave function satisfies the schr\"odinger equation, while the appropriate equation of motion of the particle is given by:
\begin{equation}
\vec{p}=m\frac{d\vec{r}}{dt}=\vec{\nabla}S
\end{equation}
where $S/\hbar$ is the phase of the wave function. One can recognize the equations of motion of particle and wave function by decomposing the wave function as $\psi=Re^{iS/\hbar}$. The result is the two equations:
\begin{equation}
\frac{\partial S}{\partial t}+\frac{|\vec{\nabla}S|^2}{2m}+V+Q=0
\end{equation}
\begin{equation}
\frac{\partial R^2}{\partial t}+\vec{\nabla}\cdot\left ( R^2\frac{\vec{\nabla}S}{m}\right )=0
\end{equation}
in which $V$ is the classical potential and $Q$ the quantum potential is defined as:
\begin{equation}
Q=-\frac{\hbar^2}{2m}\frac{\nabla^2R}{R}
\end{equation}
The first equation is Hamilton--Jacobi equation with an additional potential, the quantum potential and the second one is the continuity equation for the probability density $|\psi|^2=R^2$.

An important property of the theory is that it is critically depends on the classical configuration space chosen to represent the system. The Bohmian trajectories are highly dependent on the classical configuration space. This comes from the fact that trajectories change by canonical transformations while states transform with unitary transformations and there is not a one--to--one map between them. For simple systems like a non relativistic particle Bohmian quantum mechanics works only when one chooses $\vec{r}(t)$ as the dynamical variable. 
For such a system, if one writes the wave equation in momentum space, and extract Bohmian trajectories, the resulting path highly differs from path obtained using the position space. This latter  path is in agreement with the statistical predictions of standard quantum mechanics\cite{boh1}. Therefore Bohmian quantum mechanics in momentum space is meaningless for a non-relativistic particle.

For complicated systems like gravity, it is not clear which description is correct and thus it is worthy to investigate loop quantum cosmology in the momentum space.
\section{Causal loop quantum cosmology}
We have constructed causal loop quantum cosmology in terms of $c$ in \cite{loop}. Here we shall examine how the causal loop quantum cosmology looks like in $p$ space. To do so we set:
\begin{equation}
F(n)=R(n)e^{iS(n)/\hbar}
\end{equation}
In the Hamiltonian constraint we are dealing with $F(n\pm 1)$. Let us assume that the wave function is analytic and use the Taylor expansions:
\begin{equation}
R(n\pm 1)=\sum_{j=0}^\infty \frac{(\pm 1)^j R^{(j)}(n)}{j!}
\end{equation}
\begin{equation}
S(n\pm 1)=\sum_{j=0}^\infty \frac{(\pm 1)^j S^{(j)}(n)}{j!}
\end{equation}
in which a superscript $^{(j)}$ means the $j$-th derivative, we have:
\[
\left ( \sum_{i=0}^\infty \frac{R^{(i)}(n)}{i!}\right )\exp\left ( \frac{i}{\hbar}\sum_{j=0}^\infty \frac{S^{(j)}(n)}{j!}\right ) -\Omega R(n)\exp(iS(n)/\hbar)+
\]
\[
\left ( \sum_{i=0}^\infty \frac{(-1)^iR^{(i)}(n)}{i!}\right )\exp\left ( \frac{i}{\hbar}\sum_{j=0}^\infty \frac{(-1)^jS^{(j)}(n)}{j!}\right )=
\]
\begin{equation}
 \hbar W(n)R(n)\exp(iS(n)/\hbar)
\end{equation}
This equation can be decomposed into real and imaginary parts:
\[
\hbar W(n)+\Omega=\left ( \sum_{i=0}^\infty \frac{R^{(i)}(n)}{i!R(n)}\right )\cos\left ( \frac{1}{\hbar}\sum_{j=1}^\infty \frac{S^{(j)}(n)}{j!}\right ) +
\]
\begin{equation}
 \left ( \sum_{i=0}^\infty \frac{(-1)^iR^{(i)}(n)}{i!R(n)}\right )\cos\left ( \frac{1}{\hbar}\sum_{j=1}^\infty \frac{(-1)^jS^{(j)}(n)}{j!}\right )
\label{e1}
\end{equation}
\[
0=\left ( \sum_{i=0}^\infty \frac{R^{(i)}(n)}{i!R(n)}\right )\sin\left ( \frac{1}{\hbar}\sum_{j=1}^\infty \frac{S^{(j)}(n)}{j!}\right ) +
\]
\begin{equation}
 \left ( \sum_{i=0}^\infty \frac{(-1)^iR^{(i)}(n)}{i!R(n)}\right )\sin\left ( \frac{1}{\hbar}\sum_{j=1}^\infty \frac{(-1)^jS^{(j)}(n)}{j!}\right )
\label{e10}
\end{equation}
The first one of these two equations plays the role of a Hamilton--Jacobi equation, while the second equation is the continuity equation.

The continuity equation can be reorganized as:
\begin{equation}
R(n+1)\sin\left(\frac{S(n+1)-S(n)}{\hbar}\right )
+ R(n-1)\sin\left(\frac{S(n-1)-S(n)}{\hbar}\right )=0
\end{equation}
This can also obtained directly from the Hamiltonian constraint difference equation.

Writing the same equation for $n\rightarrow n+1$ and combining the two equations we have:
\[
R(n+2)R(n+1)\sin\left (\frac{S(n+2)-S(n-1)}{\hbar}\right )
=
\]
\begin{equation}
 R(n)R(n-1)\sin\left (\frac{S(n)-S(n-1)}{\hbar}\right )
\end{equation}
Defining
\begin{equation}
J(n)=R(n)R(n-1)\sin \left (\frac{S(n)-S(n-1)}{\hbar}\right )
\end{equation}
the continuity equation reads:
\begin{equation}
J(n+2)=J(n)
\end{equation}
which means that we have two constants:
\begin{equation}
J(n_{\textit{even}})=\textit{Constant}
\end{equation}
\begin{equation}
J(n_{\textit{odd}})=\textit{Constant}
\end{equation}
It is instructive to obtain the asymptotic limit of Hamilton--Jacobi and continuity equation. The equation (\ref{e1}) is in fact:
\[
\hbar W+\Omega=(1+\frac{R'}{R}+\frac{R''}{2R}+\cdots)\cos (\frac{S'}{\hbar}+\frac{S''}{2\hbar}+\cdots )
+
\]
\begin{equation}
(1-\frac{R'}{R}+\frac{R''}{2R}+\cdots)\cos (-\frac{S'}{\hbar}+\frac{S''}{2\hbar}+\cdots )
\end{equation}
and thus:
\begin{equation}
S'^2+\hbar^3W+\hbar^2(\Omega-2)-\frac{\hbar^2R''}{R}+\cdots=0
\label{are}
\end{equation}
Thus in the asymptotic limit Hamilton--Jacobi equation looks like the equation one expects for a continuous system. In a similar way one has for continuity equation:
\begin{equation}
(R^2S')'+\cdots=0
\end{equation}
In this continuum limit one can combine back these two equations to get a WDW--like equation:
\begin{equation}
-\Phi''+W\Phi+(\Omega-2)\Phi=0
\end{equation}
with $\Phi=Re^{iS/\hbar}$.

In order to have a complete causal interpretation of the theory we have to clarify how one can obtain the trajectories. The main problem is that equation (\ref{e1}) which plays the role of Hamilton--Jacobi equation contains higher derivatives of both $S$ and $R$. As we saw above, in the ccontinuumlimit this equation is just an ordinary Hamilton--Jacobi equation including the standard form of the quantum potential (containing only the second derivative of $R$). This means that the derivative of $S$ corresponds to the canonical momentum (the guidance relation). Writing the wavefunction as $\psi=\tilde{R}e^{i\tilde{S}/\hbar}$ and using equation (\ref{e0}) we have:
\begin{equation}
R=\textit{Constant}\times \left ( |n+\epsilon+1/4|^{3/2}- |n+\epsilon-1/4|^{3/2} \right ) \tilde{R}
\end{equation}
\begin{equation}
S=\tilde{S}-k\hbar\ell_0 n +\textit{Constant}
\end{equation}
The guidance relation is thus 
\begin{equation}
\frac{3c}{16\pi G\gamma}=\frac{\partial \tilde{S}}{\partial p}
\end{equation}
which leads to 
\begin{equation}
\frac{dS}{dn}=-\ell_0\hbar\gamma\frac{da}{dt}
\end{equation}
and using $p=\frac{\gamma}{3}\ell_p^2\ell$ (in which $\ell_p^2=8\pi G\hbar$) one can obtain $a(t)$ as:
\begin{equation}
\frac{da}{dt}=\left . -\frac{1}{\ell_0\hbar\gamma}\frac{dS}{dn}\right |_{n=3a^2/4\gamma\ell_0\ell_p^2-\epsilon}
\label{sde}
\end{equation}

It is instructive to write equation (\ref{e1}) in terms of $a$ and separating the usual Hamilton--Jacobi term $S'^2$ and other terms. A simple calculation leads to:
\begin{equation}
\left (\frac{dS}{da}\right )^2+V+{\cal Q}\left (\frac{dR}{da},\frac{d^2R}{da^2},\frac{d^3R}{da^3},\cdots ;\frac{dS}{da},\frac{d^2S}{da^2},\frac{d^3S}{da^3},\cdots\right )=0
\label{AA1}
\end{equation}
in which the \textit{classical potential} is
\begin{equation}
V=\left (\frac{3}{16\pi G}\right )^2 a^2\left (k+\frac{\hbar}{\gamma^2\ell_0^2} \right )
\label{AA2}
\end{equation}
and the \textit{quantum potential}  is given by:
\[ 
{\cal Q}=2\hbar^2-S'^2-\frac{\hbar^2}{2R}\left\{\sin\left (\frac{2S'}{\hbar}\right )\cos\left ( \frac{2}{\hbar}\sum_{k=1}^\infty\frac{S^{(2k+1)}}{(2k+1)!}\right )+\right .
\]
\[
\left . \cos\left (\frac{2S'}{\hbar}\right )\sin\left ( \frac{2}{\hbar}\sum_{k=1}^\infty\frac{S^{(2k+1)}}{(2k+1)!}\right )\right \} \times
\]
\begin{equation}
\left \{ \frac{\sum_{j=0}^\infty\frac{(-1)^jR^{(j)}}{j!}}{\sin\left ( \frac{1}{\hbar}\sum_{m=1}^\infty\frac{S^{(m)}}{m!}\right )} - \frac{\sum_{j=0}^\infty\frac{R^{(j)}}{j!}}{\sin\left ( \frac{1}{\hbar}\sum_{m=1}^\infty\frac{(-1)^mS^{(m)}}{m!}\right )}\right \}
\end{equation}
in which we have also use the continuity equation (\ref{e10}).

Although the quantum potential contains higher derivatives, one can still insists on the equation (\ref{sde}) as the guidance relation, written in terms of $a$ as:
\begin{equation}
\frac{dS}{da}=-\frac{3}{16\pi G}a\dot{a}
\label{AA4}
\end{equation}
The higher derivatives of $S$ are thus related to acceleration and higher time derivatives of $a$, for example $\frac{d^2S}{da^2}=-\frac{3}{16\pi G}\left ( \frac{a\ddot{a}}{\dot{a}}+\dot{a}\right )$. The quantum potential then can be written as:
\[
{\cal Q}=\left ( 1-\frac{\ell_0^2\gamma^2}{2}\dot{a}^2\right )\left (-\hbar^2\frac{a}{R}\frac{d}{da}\left (\frac{1}{a}\frac{dR}{da}\right )\right ) +
\]
\begin{equation}
 \frac{3\hbar^2}{2}\frac{a\ddot{a}}{\dot{a}^2}\left (1-\frac{5}{6}\ell_0^2\gamma^2\dot{a}^2\right )\left \{ \frac{2}{3aR}\frac{dR}{da}+\frac{\gamma^2\ell_0^2\ell_p^4}{81aR}\frac{d}{da} \left ( \frac{1}{a}\frac{d}{da}\left ( \frac{1}{a}\frac{dR}{da}\right ) \right ) \right \}+\cdots
\label{AA3}
\end{equation}
The quantum potential can be either viewed as a function of $a$, $\dot{a}$, $\ddot{a}$, $\cdots$, or by virtue of the guidance relation as a function of $a$ only. 

Equations (\ref{AA1}), (\ref{AA2}), (\ref{AA3}), (\ref{AA4}), and the wave equation (\ref{AA5}) are the essential relations of the theory. They determine a unique Bohmian trajectory. To see this note that the wave equation (\ref{AA5}) has a unique solution provided that necessary initial conditions are specified, and on the other hand the guidance relation (\ref{AA4}) is a  first order differential equation which is always integrable. This is why one should insist on the guidance relation of this form and include all other terms in the definition of the quantum potential.

In order to justify this conclusion more, let's to differentiate the Hamilton--Jacobi equation (\ref{AA1}) with respect to $a$ which gives us the equation of motion of $a$. On using the guidance relation, the result is:
\begin{equation}
\frac{d}{dt}(a\dot{a})=-\frac{1}{2a}\left (\frac{3}{16\pi G}\right )^2\frac{d}{da}(V+{\cal Q})
\label{AA6}
\end{equation}
For an empty universe with a cosmological constant $\Lambda$ we have 
\begin{equation}
H_m(a)=\frac{1}{8}\frac{\Lambda a^3}{16\pi G}
\end{equation}
and thus 
\[
\hbar W= -\frac{8\pi G}{3}\frac{\gamma^3\ell_0^3\hbar\Lambda a^3}{\left |a^2+8\pi G\gamma\hbar\ell_0/3\right |^{3/2}-\left |a^2-8\pi G\gamma\hbar\ell_0/3\right |^{3/2}}
\]
\begin{equation}
\equiv
-\frac{1}{3}\gamma^2\ell_0^2\Lambda_{eff}(a) a^2
\end{equation}
Therefore the Hamilton--Jacobi equation (\ref{AA1}) and the equation of motion (\ref{AA6}) read as:
\begin{equation}
\frac{\dot{a}^2}{a^2}+\frac{k}{a^2}-\frac{\Lambda_{eff}(a)}{3}+\left (\frac{16\pi G}{3}\right )^2\frac{{\cal Q}}{a^4}=0
\end{equation}
\begin{equation}
\frac{\ddot{a}}{a}+\frac{\dot{a}^2}{a^2}+\frac{k}{a^2}-\frac{2\Lambda_{eff}(a)}{3}
-\frac{a}{6}\frac{d\Lambda_{eff}(a)}{da}
+\left (\frac{16\pi G}{3}\right )^2\frac{d{\cal Q}/da}{2a^3}=0
\end{equation}
which are just the classical Friedmann equations corrected by the quantum potential (${\cal Q}$) and the quantum force ($d{\cal Q}/da$) and also by discreteness of space corrections to the cosmological constant. This latter one is not present for large scale factors.
\section{Bohmian trajectories}
\begin{figure}[htb]
\begin{center}
\epsffile{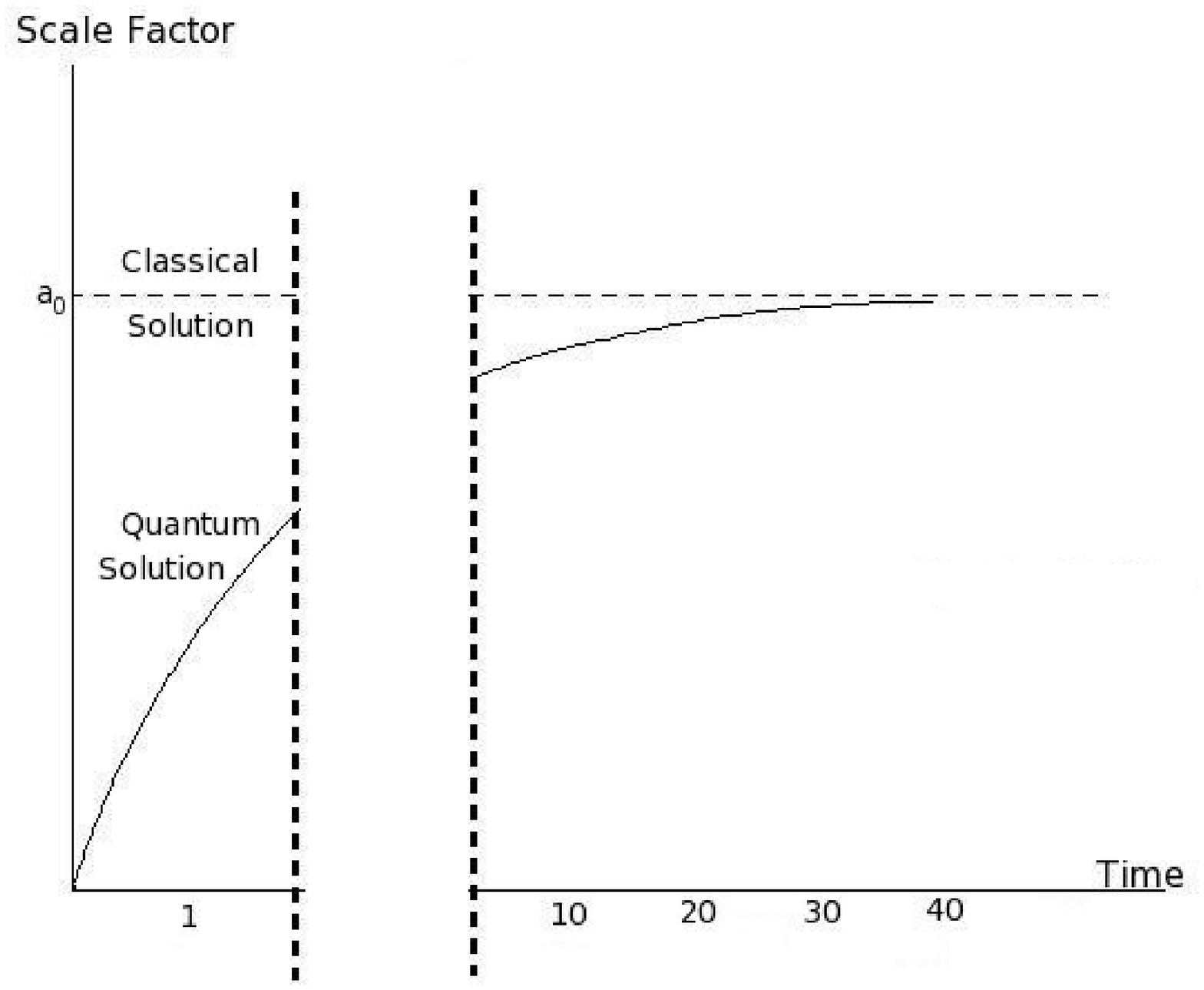}
\end{center}
\caption{Bohmian trajectory for a flat source-less universe. The scale factor is normalized to Planck length and time is normalized to Planck time.}
\label{f.1}
\end{figure}

In order to show how the Bohmian trajectories could be, we first apply it to the solution of loop quantum cosmology without any source, i.e $W=0$ and for the flat case. Using the solutions obtained in \cite{loop} and choosing the following linear combination:
\begin{equation}
\psi(\ell)=\left (\frac{6}{\gamma}\right )^{3/2}\ell_p^{-3} \frac{1}{|\ell+\ell_0|^{3/2}-|\ell-\ell_0|^{3/2}} (1+\alpha e^{i\phi}\ell)
\label{wa}
\end{equation}
where $\alpha$ and $\phi$ are constants. On can obtain:
\begin{equation}
\frac{dS}{dn}= \frac{4\hbar\ell_0\alpha\sin\phi}{1+8\alpha\ell_0\cos\phi (n+\epsilon)+16\alpha^2\ell_0^2(n+\epsilon)^2}
\end{equation}
Using the relation (\ref{sde}), the corresponding Bohmian trajectory is plotted in figure (\ref{f.1}).  The classical trajectory of such a universe is a constant scale factor plotted as a dotted line.  As it can be seen in the figure, the Bohmian trajectory for such a system converges to the classical path after several times the Planck time. The free parameter $\alpha$ in the wave function determines the value of $a_0$.  At times of order or less than the Planck time, Bohmian trajectory highly differs from the classical one. The quantum potential is responsible for this behavior. At first this seems very strange, as the trajectories indicate that the quantum analog of Minkowfski space starts from a big bang. This only shows that it is possible to choose the wave function (equation (\ref{wa})) such that this happens. Therefore Bohm theory admits us to interpret the Minkowski space as an expanding universe with a high expansion rate, so that after passing several times of Planck time we arrive at Minkowski space.

\epsfxsize=5in

\begin{figure}[htb]
\begin{center}
\epsffile{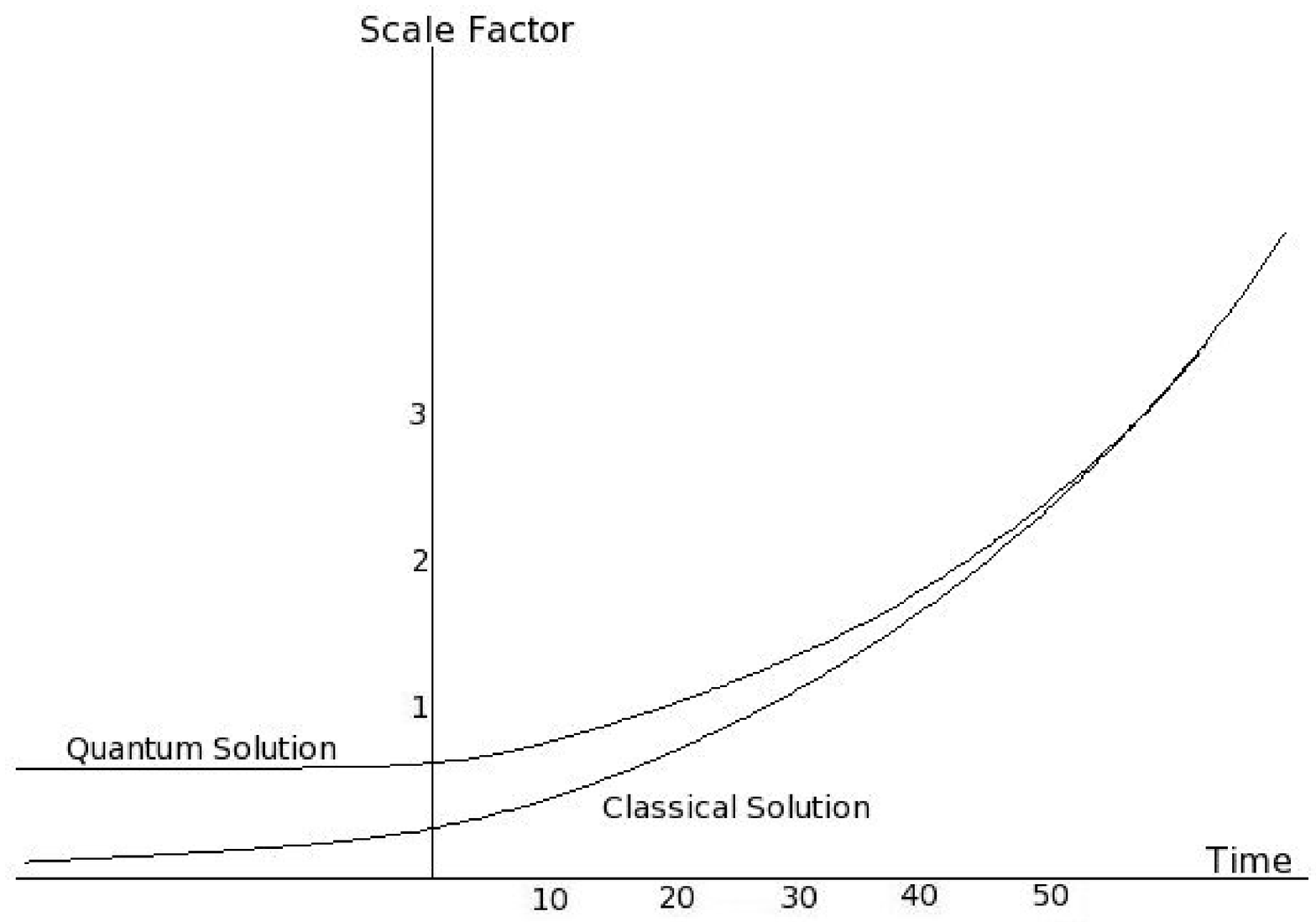}
\end{center}
\caption{Bohmian trajectory for a flat matter-less universe with a small cosmological constant. The scale factor is normalized to Planck length and time is normalized to Planck time.}
\label{f.2}
\end{figure}

As a second example we consider the flat case with a small cosmological constant
and without matter fields. We can use the solutions obtained in \cite{var}. Using the appropriate linear combination as (first solution) $+ i$ (second solution), one can obtain a Bohmian trajectory with correct classical limit, while one avoids the initial singularity. This trajectory is plotted in figure (\ref{f.2}). Again Bohmian quantum corrections are negligible at classical times, while below the Planck time quantum potential causes the trajectory to differ from the classical trajectory slightly.

\epsfxsize=5in

\textbf{Acknowledgments}
This work is partly supported by a grant from University of Tehran and partly by a grant from the center of excellence in the structure of matter of Department of Physics of University of Tehran.

\end{document}